\documentclass[twocolumn,showpacs,preprintnumbers,amsmath,amssymb]{revtex4}
\usepackage{graphicx}
\usepackage{dcolumn}
\usepackage{bm}

\newcommand{\be}{\begin{equation}}
\newcommand{\ee}{\end{equation}}
\newcommand{\bea}{\begin{eqnarray}}
\newcommand{\eea}{\end{eqnarray}}
\newcommand{\mbb}{\mathbb}
\newcommand{\ti}{\times}
\newcommand{\half}{\frac{1}{2}}

\newcommand{\mc}{\mathcal}

\begin{document}
\preprint{DAMTP-2006-55}
\preprint{hep-ph/0607138}
\title{Seeing the Invisible Axion in the Sparticle Spectrum}
\author{Joseph P. Conlon}
\affiliation{%
DAMTP, Centre for Mathematical Sciences, Wilberforce Road, Cambridge
CB3 0WA, UK}%
\begin{abstract}
I describe how under favourable circumstances the invisible axion may manifest its existence at
the LHC through the sparticle spectrum; in particular through a gluino
$\sim \ln (M_P/m_{3/2})$ times heavier than other gauginos.
\end{abstract}
\pacs{11.25.Mj 11.25.Wx 12.10.-g 14.80.Ly 14.80.Mz}
\maketitle

The invisible axion is the best-motivated solution to the strong CP
problem. The QCD Lagrangian in principle could contain a CP-violating
$\theta$ angle,
\be
S = -\frac{1}{4g^2} \int d^4x F_{\mu \nu}^a F^{a,\mu \nu}
+ \ldots +\frac{\theta}{32 \pi^2} \int d^4 x F^a_{\mu \nu} \tilde{F}^{a,\mu
  \nu}.
\ee
The presence of such a $\theta$-angle would give rise to observable CP
violation in strong interactions and in particular to an electric
dipole moment for the neutron. Experiment constrains $\vert \theta
\vert \lesssim 10^{-10}$ \cite{Harris, Romalis}. As
$\theta$ is periodic with an allowed range $-\pi < \theta < \pi$, this
is unnatural.
The problem of why $\vert \theta \vert$ is so small is the
strong CP problem.

In the Peccei-Quinn solution \cite{PecceiQuinn}, $\theta$ is promoted to a dynamical
 field $\theta(x)$.
\be
\mc{L}_\theta = \mc{L}_{QCD} - \frac{1}{2} f_a^2 \partial_\mu \theta
\partial^\mu \theta + \frac{\theta}{32 \pi^2} F^a_{\mu \nu} \tilde{F}^{a,\mu
  \nu}.
\ee
The dimensionful quantity $f_a$ is known as the axion decay constant.
The action for the canonically normalised field $a \equiv f_a \theta$ is
\be
\mc{L} = -\frac{1}{2}  \partial_\mu a \partial^\mu a + \frac{a}{32
  \pi^2 f_a}  F^a_{\mu \nu} \tilde{F}^{a,\mu \nu}.
\ee
QCD instanton effects
generate a potential for $\theta$. This potential may be computed
 using the pion Lagrangian (see e.g. section 23.6 of \cite{Weinberg}), giving
\be
\label{axionpotential}
V(a) \sim m_\pi^2 f_\pi^2 \left( 1 - \cos \left( \frac{a}{f_a} \right)
 \right),
\ee where $f_\pi \sim 90 \hbox{MeV}$ is the pion decay constant.
 The potential (\ref{axionpotential}) dynamically
minimises $\theta$ at zero, thus solving the strong CP problem.

$a$ is known as the (invisible) axion. It is extremely light and has
mass
\be
\label{axionmass}
m_a \sim \frac{m_\pi^2 f_\pi^2}{f_a^2} \sim \left(\frac{10^9
  \hbox{GeV}}{f_a}\right) 10^{-2} \hbox{eV}.
\ee
Astrophysical and cosmological constraints
imply that if $a$ exists, the decay constant $f_a$ lies within a narrow window
$10^9 \hbox{GeV} \lesssim f_a \lesssim 3 \ti 10^{11} \hbox{GeV}$ \cite{hepph0509198}.
The lower bound is hard and comes from supernova cooling. The upper
bound assumes a standard cosmology and comes from the requirement that
the energy density in the axion field due to primordial misalignment does not exceed the
current dark matter density. This may however be relaxed with non-standard
cosmologies, for example involving late-time inflation.

Although a solution of the strong CP problem only requires the axion to couple to
QCD, in most axion models the axion also couples to QED. In this case
axions can convert to photons in a background magnetic field through the
Primakoff effect. This $a \gamma \gamma$ coupling
is the basis for direct axion searches such as axion helioscopes
\cite{hepex0411033}, cavity experiments \cite{astroph0603108} or
laser-based searches \cite{hepex0507107}.
There is no unambiguous detection of an axion in any of these
experiments.
There does exist a possible signal in the PVLAS experiment
\cite{hepex0507107} which
is however in tension with existing astrophysical bounds.

The purpose of this note is to point out that under favourable
assumptions the invisible axion may also indirectly manifest its existence
in particle colliders such as the LHC, through the sparticle
spectrum and in particular through the existence of a gluino very
much heavier than other gauginos by a factor $\sim \ln (M_P/m_{3/2})
\sim 30$. This will follow from considerations of moduli stabilisation
in string and supergravity scenarios.

String theory remains the best candidate for the ultraviolet physics underlying
the Standard Model. Axions in string theory are both natural and
abundant. In braneworld scenarios, they descend from higher-dimensional components of
antisymmetric RR-form fields. The
Dp-brane action is
\bea
\label{braneaction}
S_{DBI} + S_{CS} & = & -\frac{2 \pi}{l_s^{p+1}} \int d^{p+1}x \, e^{-\phi} \sqrt{g +
  B + 2 \pi \alpha' F} \nonumber \\
& + & i \frac{2 \pi}{l_s^{p+1}} \int e^{B + 2\pi \alpha'
  F} \wedge \sum C_q,
\eea
with $l_s = 2 \pi \sqrt{\alpha'}$.
Expanding the Chern-Simons part of (\ref{braneaction}) gives a term
$$
\frac{1}{2 (2 \pi)l_s^{p-3}} \int_{\mbb{M}_4 \ti \Sigma} F \wedge F \wedge C_{p-3}.
$$
$\Sigma$ is the $(p-3)$-dimensional Calabi-Yau cycle wrapped by the brane.
 The relevant axionic coupling is generated by the
appropriate component of the RR potential $C_{p-3}$.

The low-energy limit of string theory is 4-dimensional $\mc{N} =1$
supergravity. In this limit, axions $a_i$ appear as the imaginary
parts of scalar components of moduli multiplets, \be a_i =
\hbox{Im}(T_i), \qquad T_i = \tau_i + i a_i, \ee where the real
parts $\tau_i$ typically parametrise the geometry of the
compactification manifold. Axions are $U(1)$-valued and thus have an
exact symmetry $a_i \to a_i + 2\pi$. The continuous shift symmetry
$a_i \to a_i + \epsilon$ is valid perturbatively, implying that up
to non-perturbative effects neither the K\"ahler potential nor
superpotential can depend on the axions. If $\Phi_j$ denote the
superfields with no axionic components, this implies that
perturbatively \bea
W(T_i, \Phi_j) & = & W(\Phi_j), \nonumber \\
\mc{K}(T_i, \Phi_j) & = & \mc{K}(T_i + \bar{T}_i, \Phi_j).
\eea

Gauge groups are realised by wrapping stacks of branes on cycles. The gauge coupling
is determined by the size of the cycle: by expanding (\ref{braneaction}), large cycles imply
weak coupling. The gauge kinetic function is the modulus associated
with the cycle wrapped by the brane.

We suppose such a stringy invisible axion does indeed exist and solves the strong
CP problem. In that case,
\be
\label{qcdaxion}
a = \hbox{Im}(T_{QCD}),
\ee
where $T_{QCD} = \tau_Q + i a$ is whatever modulus is the QCD gauge
kinetic function.
This follows from the supergravity couplings \cite{WessBagger},
\be
\mc{L} \sim -\frac{\hbox{Re}(f_a)}{4} F_{\mu \nu}^a F^{a,\mu \nu}
+i \frac{\hbox{Im}(f_a)}{8} F^a_{\mu \nu} \tilde{F}^{a, \mu
  \nu} + \ldots.
\ee
Depending on exactly how the moduli are stabilised
the physical axion may be an admixture of (\ref{qcdaxion}) with other moduli.
The saxion $\tau_Q$ is an ordinary scalar field and thus will have
non-derivative couplings to matter. If $\tau_Q$ were massless,
these couplings would
generate long-range Yukawa forces. The non-observation of such fifth
forces implies this cannot hold - the saxion must be massive and a potential must be
generated for it.

The generation of moduli potentials in string theory - i.e. moduli
stabilisation - has received much recent
attention \cite{hepth0105097, hepth0301240}. The supergravity F-term
potential is determined by both the K\"ahler potential and superpotential,
\be
\label{ftermpot}
V_F = e^{\mc{K}} \left( \mc{K}^{i \bar{j}} D_i W D_{\bar{j}} \bar{W}
-3 \vert W \vert^2 \right).
\ee
As noted above, the axion shift symmetry implies the modulus $T_{QCD}$ cannot appear perturbatively in the
superpotential. $T_{QCD}$ can appear non-perturbatively and
nonperturbative superpotentials represent a popular approach to
moduli stabilisation.
For example, in the KKLT scenario \cite{hepth0301240},
the K\"ahler and superpotential are given by
\bea
\label{kkltkw}
\mc{K} & = & -2 \ln (\mc{V}(T_i + \bar{T}_i)), \nonumber \\
W & = & W_0 + \sum_i A_i e^{-a_i T_i}.
\eea
The constant $W_0$ comes from 3-form fluxes while $\mc{V}$ is the
volume of the internal space.
The moduli are stabilised by solving $D_i W = 0$ for all K\"ahler
moduli. A further strong justification for the presence of nonperturbative superpotentials is
that they can naturally generate the weak scale/Planck scale hierarchy, as in e.g. the
racetrack scenario \cite{racetrack} (see \cite{hepth0606262} for some
recent work) or the exponentially large volume models of
\cite{hepth0502058, hepth0505076}.

However, for the axion $a$ to solve the strong CP problem
its saxion partner $\tau_{Q}$ should \emph{not} be stabilised
through a non-perturbative superpotential. The axionic solution to
the strong CP problem requires that QCD instantons dominate the axion
potential. These generate a potential (\ref{axionpotential}) of magnitude
$\sim \Lambda_{QCD}^4 \sim 10^{-75} M_P^4$.
However a superpotential term
\be
\label{qcdnpw}
W = W_0 + \ldots + A_Q  e^{-a T_{QCD}} + \ldots \, ,
\ee
as may appear in (\ref{kkltkw}),
depends on the axion
$a$ through the phase of the exponent and thus generates an axion mass
through the potential (\ref{ftermpot}).
The typical size of this potential is $\sim e^{\mc{K}} W
\bar{W} \sim m_{3/2}^2 M_P^2$, which for a TeV-scale gravitino mass is
$\sim (10^{11} \hbox{GeV})^4$. The putative QCD axion $a$ obtains a
mass at a scale comparable to $\tau_Q$, $m_a \sim m_{\tau_Q} \sim m_{3/2} \sim 1 \hbox{TeV}$. This is much
larger than the range (\ref{axionmass}) $10^{-6}
\hbox{eV} \lesssim m_a \lesssim 10^{-3} \hbox{eV}$ associated with QCD
effects in the allowed $f_a$ window. The term (\ref{qcdnpw}) washes out the
effect of QCD instantons and the
minimum for $a$ becomes uncorrelated with the vanishing of the
$\theta$-angle, leaving the strong CP problem no longer solved.
Consequently $\tau_Q$ must instead be stabilised perturbatively through the
structure of the K\"ahler potential - as $\mc{K} = \mc{K}(T + \bar{T})$, this does not
generate a potential for the axion.

However, the above argument do not apply
to the moduli $T_{SU(2)}$ and $T_{U(1)}$ controlling the $SU(2) \ti
U(1)$ gauge couplings - here the instanton amplitudes are so small the
$\theta$ angle is irrelevant. These moduli must be stabilised to avoid
fifth forces, but there is no restriction on how this is achieved.
The favourable assumption we make is that at least one of these
moduli is stabilised nonperturbatively, while $T_{QCD}$ is
stabilised perturbatively. This assumption requires
intersecting-brane rather than GUT phenomenology, as in the latter case
only a single modulus controls all gauge couplings and the above
assumption does not make sense.

Very generally,
the pattern of the MSSM soft terms is determined by how supersymmetry
is broken. We assume gravity-mediation - in this case,
the soft terms are determined by the
structure of the hidden sector - i.e. moduli - potential. For a gauge group with gauge kinetic function $f_a$,
the gaugino masses are given by
\be
M_a = \half \frac{1}{\hbox{Re } f_a} \sum_\alpha F^\alpha \partial_\alpha f_a.
\ee
The F-terms $F^\alpha$ are defined by
\bea
\label{ftermformula}
F^\alpha & = & e^{\mc{K}/2} \sum_{\bar{\beta}} \mc{K}^{\alpha \bar{\beta}} D_{\bar{\beta}}
\bar{W} \\
& = & e^{\mc{K}/2} \sum_{\bar{\beta}} \mc{K}^{\alpha \bar{\beta}} \partial_{\bar{\beta}}
\bar{W} + e^{\mc{K}/2} \sum_{\bar{\beta}}  \mc{K}^{\alpha \bar{\beta}}
(\partial_{\bar{\beta}} \mc{K}) \bar{W} \nonumber.
\eea
In the case that $f_a = T_a$, the associated gaugino
mass is
\be
\label{GauginoMass}
M_a = \frac{F^a}{2 \tau_a}.
\ee
Note the mass in (\ref{GauginoMass}) is a Lagrangian parameter and applies at the compactification scale; to determine the
physical mass we must run this down to the \hbox{TeV} scale.

The point of the favourable assumption is that
if a modulus $T_a$ is stabilised through non-perturbative
superpotential corrections, the associated F-term
is generically suppressed. To be explicit, in an expansion in
$\frac{1}{\ln(M_P/m_{3/2})}$,
 the two contributions to (\ref{ftermformula})
cancel to leading order \cite{hepth0605141} (see also \cite{hepth0411066}).
The magnitude of the resulting F-term is then
\be
F^a \sim \frac{2 \tau_a m_{3/2}}{\ln (M_P/ m_{3/2})}.
\ee
and the gaugino associated to $T_a$ has a mass suppression,
\be
\label{GauginoSuppression}
M_a \sim \frac{m_{3/2}}{\ln(M_P/m_{3/2})}.
\ee
This logarithmic suppression can be shown to be entirely a feature of the non-perturbative
 stabilisation. In particular, it does not depend on whether the stabilisation is
 approximately supersymmetric or not.

We argued above that $\tau_Q = \hbox{Re}(T_{QCD})$
should be stabilised perturbatively through the structure of
the K\"ahler potential.
The K\"ahler potential is non-holomorphic and
 thus hard to compute and so it is difficult to be explicit. However
the only fact we need is that for this case the logarithmic
suppression of (\ref{GauginoSuppression}) does \emph{not}
hold. Instead we obtain the generic behaviour
\be
M_a \sim m_{3/2}.
\ee
Such behaviour is indeed found in explicit models of perturbative
stabilisation (e.g. see \cite{hepph0511162}).

We now come to the point of this letter.
What the Peccei-Quinn solution to the strong CP problem tells us is that
whatever the modulus controlling the QCD gauge coupling is, it
should be stabilised perturbatively through the K\"ahler potential
rather than non-perturbatively through the
superpotential. This is to avoid generating a potential for the QCD axion.
In this case the associated
gaugino - i.e. the gluino - will not have a suppressed mass, and we expect
$m_{\tilde{g}} \sim m_{3/2}$ at the compactification scale.

However under the favourable circumstances above, at least one of the moduli
$T_{SU(2)}$ and $T_{U(1)}$ is stabilised nonperturbatively and its
associated gaugino mass will be suppressed.
Consequently either $m_{\tilde{B}}$ or $m_{\tilde{W}}$ will
have a suppressed mass,
\be
\label{gauginosupp}
M_a \sim \frac{m_{3/2}}{\ln(M_P/m_{3/2})}.
\ee
If (\ref{gauginosupp}) holds,
anomaly-mediated contributions are also relevant for the exact gaugino masses:
it is a numerical coincidence that $\ln (M_P/1 \hbox{TeV}) \sim 0.4 (8 \pi^2)$.

There is then a small hierarchy between the masses of the gluino
and the lightest gaugino set by $\ln (M_P/m_{3/2})$.
The axionic solution to the strong CP probem
therefore suggests a distinctive gaugino spectrum, with for example
\be
\label{heavygluino}
m_{\tilde{g}} \sim 30 m_{\tilde{W}}.
\ee
As the gluino mass tends to increase under RG flow this hierarchy
will not be diluted by the running to low energy.
Such a hierarchy may be unnatural from the viewpoint
of low-energy field theory. However the relations (\ref{gauginosupp})
and (\ref{heavygluino}) are another reminder that
naturalness in string theory and naturalness in effective field theory
are cognate but non-identical concepts.

The spectrum of (\ref{gauginosupp}) is very distinctive and can be easily distinguished
from MSUGRA models, where the gaugino mass ratios are universal at the compactification scale and
the physical masses have $m_{\tilde{g}} \sim 5 m_{\tilde{W}}$.
This justifies our original claim: the invisible axion could manifest
its existence at particle colliders such as the LHC through a
very distinctive gaugino spectrum.

Let us compare the above approach to `seeing' the axion
to more conventional routes. One advantage is that it implies the
existence of a QCD axion could have consequences for areas
such as collider physics very different from the astrophysical topics
normally considered. The above gaugino spectrum is also very distinctive,
involving a sharp hierarchy between the lightest and heaviest
gauginos.

Another advantage is that the arguments above are insensitive to the value of the decay constant
$f_a$ and whether or not the axion couples to photons. While decay constants
$f_a \gtrsim 10^{12} \hbox{GeV}$ are disfavoured by standard
cosmology, this upper bound is not solid and may be evaded by
non-standard cosmologies or small initial axion misalignment,
$\vert \theta_i \vert \ll 1$. Furthermore, `generic' string compactifications give
$f_a \sim 10^{16} \hbox{GeV}$ - for recent discussions of $f_a$ in
string compactifications see \cite{hepth0602233, hepth0605206, hepth0605256}.
In the case that $f_a \gg 10^{12} \hbox{GeV}$ or
$g_{a \gamma \gamma} = 0$, detection by direct search experiments would be extremely
difficult. However the above arguments are unaltered and the gluino
hierarchy may still be seen. Other than the vanishing of $\theta_{QCD}$, the relation
(\ref{heavygluino}) could then be the only way the existence of the
axion is manifested experimentally.

An obvious disadvantage of the above approach is that it is indirect:
even if seen, the gaugino hierarchy of (\ref{heavygluino}) would be at best
an indication of an axion rather than a detection. It also requires a
favourable assumption. In principle all  moduli could be stabilised
perturbatively, in which case there exists quasi-universal behaviour
$M_a \sim m_{3/2}$ with no distinct gluino hierarchy: a failure to
observe the hierarchy (\ref{heavygluino}) would not rule out the invisible axion.

We also note it would be straining the LHC to actually see both the
light gaugino and the hierarchically heavier gluino. Doing so would
require a favourable sparticle spectrum, with a
very light gaugino at $\sim 100 \hbox{GeV}$ near the LEP bounds
and a gluino at the extreme reach of the LHC with $m_{\tilde{g}} \sim
2 \hbox{TeV}$. As $m_{\tilde{g}} \sim m_{3/2}$, this would also
suggest a generally heavy scalar spectrum with TeV scalars having $m_i
\sim m_{3/2}$. In this case
the only sparticles accessible at the LHC may be winos or binos.

Let us conclude by restating the assumptions and result. We assume gravity-mediated
supersymmetry breaking. To avoid generating a potential for the QCD axion, the modulus
$\tau_{QCD}$ controlling the QCD gauge coupling must be stabilised perturbatively.
We also assume one of the moduli $\tau_{SU(2)}$ or $\tau_{U(1)}$ is stabilised through a
nonperturbative superpotential. In this case there is a generic hierarchy of $\ln(M_P/m_{3/2})$ between the
gluino and the lightest gaugino.

We find it both interesting and amusing that the invisible
axion may manifest its existence as above in the perhaps unlikely spot of the
sparticle spectrum. This may be particularly important if the
properties of the axion are such as to make its direct detection impossible.

{\noindent \bf Acknowledgements}

I am grateful to EPSRC for a studentship and Trinity College,
Cambridge for a research fellowship. I thank F. Quevedo for reading a draft of the manuscript and the
ICTP `String Vacua and the Landscape' workshop for hospitality.

\bibliography{apssamp}

\end{document}